\documentclass[a4paper,usenatbib]{mn2e}
\usepackage{journals}
\usepackage{natbib}

\usepackage[T1]{fontenc}
\usepackage{ae,aecompl}
\usepackage{threeparttable}

\newcommand{\EQ}[1] {equation~(\ref{#1})}

\newcommand{\FIG}[1] {Figure~\ref{#1}}
\newcommand{\TAB}[1] {Table~\ref{#1}}

\newcommand{\VEC}[1] {{\boldsymbol{{ #1}}}}
\newcommand{\MX}[1] {{\mathbfss{{#1}}}}

\newcommand{\lnm}{\textsc{linimoss}}

\usepackage{graphicx}	
\usepackage{amsmath}	
\usepackage{amssymb}	
\usepackage{color}

\title[PTAs with a dynamical Solar-system model]{Studying the Solar system dynamics using pulsar timing arrays and the LINIMOSS dynamical model}

\author[Y. J. Guo et al.]{
Y. J. Guo$^{1}$, 
G. Y. Li$^{2}$\thanks{Deceased},
K. J. Lee$^{1,3}$\thanks{E-mail: kjlee@pku.edu.cn},
R. N. Caballero$^{1}$
\\
$^{1}$Kavli Institute for Astronomy and Astrophysics, Peking University, Beijing 100871, China\\
$^{2}$Purple Mountain Observatory, Academia Sinica, Nanjing 210008, China\\
$^{3}$National Astronomical Observatories, Chinese Academy of Sciences, Beijing 100012, China\\
}

\date{Accepted XXX. Received YYY; in original form ZZZ}

\pubyear{2019}

\begin{document}
\label{firstpage}
\pagerange{\pageref{firstpage}--\pageref{lastpage}}
\maketitle

\begin{abstract}
Pulsar timing arrays (PTAs) can be used to study the Solar-system ephemeris (SSE), the errors of 
which can lead to correlated timing residuals 
and significantly contribute to the PTA noise budget.
Most Solar-system studies with PTAs assume the dominance of the 
term from the shift of the Solar-system barycentre (SSB).
However, it is unclear to which extent this approximation can be valid, 
since the perturbations on the planetary 
orbits may become important as data precision keeps increasing.
To better understand the effects of SSE uncertainties on pulsar timing, 
we develop the \lnm{} dynamical model of the Solar system, based on the SSE of Guangyu Li.
Using the same input parameters as DE435, 
the calculated planetary positions by \lnm{} are compatible with DE435 at 
centimetre level over a 20-year timespan, 
which is sufficiently precise for pulsar-timing applications.
We utilize \lnm{} to investigate the effects of SSE errors on pulsar timing in a fully dynamical way,
by perturbing one SSE parameter per trial and examining the induced timing residuals.
For the outer planets, the timing residuals are dominated by the SSB shift, 
as assumed in previous work.
For the inner planets, the variations in the orbit of the Earth are more prominent,
making previously adopted assumptions insufficient.
The power spectra of the timing residuals have complex structures, which may introduce false signals in 
the search of gravitational waves.
We also study how to infer the SSE parameters using PTAs, 
and calculate the accuracy of parameter estimation.

\end{abstract}

\begin{keywords}
	pulsar:general -- methods: data analysis
\end{keywords}

\section{Introduction}

Timing observations of multiple pulsars can be used to study the phenomena which affect all pulsars and induce spatially-correlated signals.
Such an ensemble of pulsars located at different sky positions is referred to as a pulsar timing array~\citep[PTA;][]{FB90}, 
primarily used in attempts for direct detection of low-frequency gravitational waves~\citep[GW;][]{HD83}.
The pulse times of arrival (TOAs) recorded at the observatory need to 
be transferred to a quasi-inertial frame located at the Solar-system 
barycentre (SSB), i.e. we need to calculate the equivalent TOA for the 
same pulse wavefront at the SSB.
This step requires the use of a Solar-system ephemeris (SSE),
which uses information on the masses and orbits of the 
Solar-system objects and provides
the position of the Earth relative to the SSB.
Possible errors in the SSE will induce
additional timing residuals which exhibit
dipolar spatial correlations between pulsars. 
This provides a way of constraining the SSE parameters, such as planetary masses, using PTAs~\citep{Champion10, CGL18}.
At the level of contemporary pulsar timing precision,
the timing signals from such SSE errors can make important contribution to the PTA noise budget~\citep{Caballero18,Lazio18}, 
and the GW upper limits and detection statistics are affected by the choice of SSEs~\citep{thk+16,Taylor17, NG18}.

Several approaches to model the SSE errors have been proposed, either to constrain the SSE parameters or mitigate the SSE errors.
\citet{Champion10} constrained the mass errors of the known planets, assuming that the mass errors are small enough
such that perturbations in the planetary orbits are negligible, and thus keeping the orbits fixed to the input data from the SSE.
Using this approximation,
we can also search for possible unmodelled objects in Keplerian, unperturbed orbits around the Sun
and put upper limits on their masses~\citep*{GLC18}.
\citet{Deng13} searched for non-physical, generic error vectors of the SSB position.
\citet{NG18} employed a physical model of SSE errors to constrain the strain amplitude of 
a nHz GW background, which includes uncertainties in the outer planets' masses, Jupiter's orbit, and coordinate-frame drifts.

All studies mentioned above model the effects of SSE errors by assuming that the changes of planetary mass or orbit only lead to linear shifts in the SSB position.
However, it has been unclear to which
extent this approach remains a valid approximation, 
since the interactions between planets can be important in high-precision pulsar-timing data.
The change in the planetary mass or orbit will also influence the orbit of the Earth,
which could be comparable to, or stronger 
than the SSB shift, if the planet is close to the Earth.
While specific tests have been previously reported 
\citep[e.g. secondary effects by deviations in the mass of Jupiter repoted in][]{Champion10},
in this work we present a detailed examination of the effects additional to linear SSB shifts
caused by possible errors in planetary masses and orbital elements for all the major planetary systems.

In order to study the effects of SSE errors on high-precision pulsar-timing data, appropriate dynamical modelling of the Solar system is needed to accurately calculate the change of planetary orbits when perturbing the SSE parameters.  
Ephemerides published by the Jet Propulsion Laboratory (JPL)
are most widely used in pulsar timing~\citep{Standish98}.
There are also the EPM and INPOP 
families of ephemerides ~\citep{Pitjeva05, Fienga08}, developed independently
by the Institute of Applied Astronomy of Saint-Petersburg 
and the Institut de M\'ecanique C\'eleste et de Calcul des \'Eph\'em\'erides, respectively,
which use different dynamical models and input data.
We take a recent ephemeris from JPL,
namely DE435 ~\citep{Folkner16}, as reference for the development of our dynamical modelling
that we use to study the effects of possible errors in SSE parameters.
Based on the PMOE planetary ephemeris of Guangyu Li (see next section), we have implemented a dynamical model similar to DE435, which describes the physics governing the motion of Solar-system bodies (to present observational accuracy),
named \emph{LI's Numerically Integrated MOdel of the Solar System} (\lnm{}). 
We use the same input parameters as DE435 
to reproduce it for demonstration of the validity of the \lnm{} model.
The dynamical model can help us to better understand the effects of perturbing SSE parameters in pulsar timing and to further
improve the methodology used to measure or constrain relevant SSE parameters.

The rest of the paper is organized as follows.
We briefly introduce the \lnm{} dynamical model in Section~\ref{sec:sse}.
The timing residuals caused by perturbing the SSE parameters are shown in Section~\ref{sec:lsaeq}, 
as well as constraints on the SSE parameters using pulsar timing.
Conclusions are made in Section~\ref{sec:end}.


\section{Dynamical model of the Solar system}
\label{sec:sse}

\begin{figure*}
\begin{center}
\includegraphics[width=1.9\columnwidth]{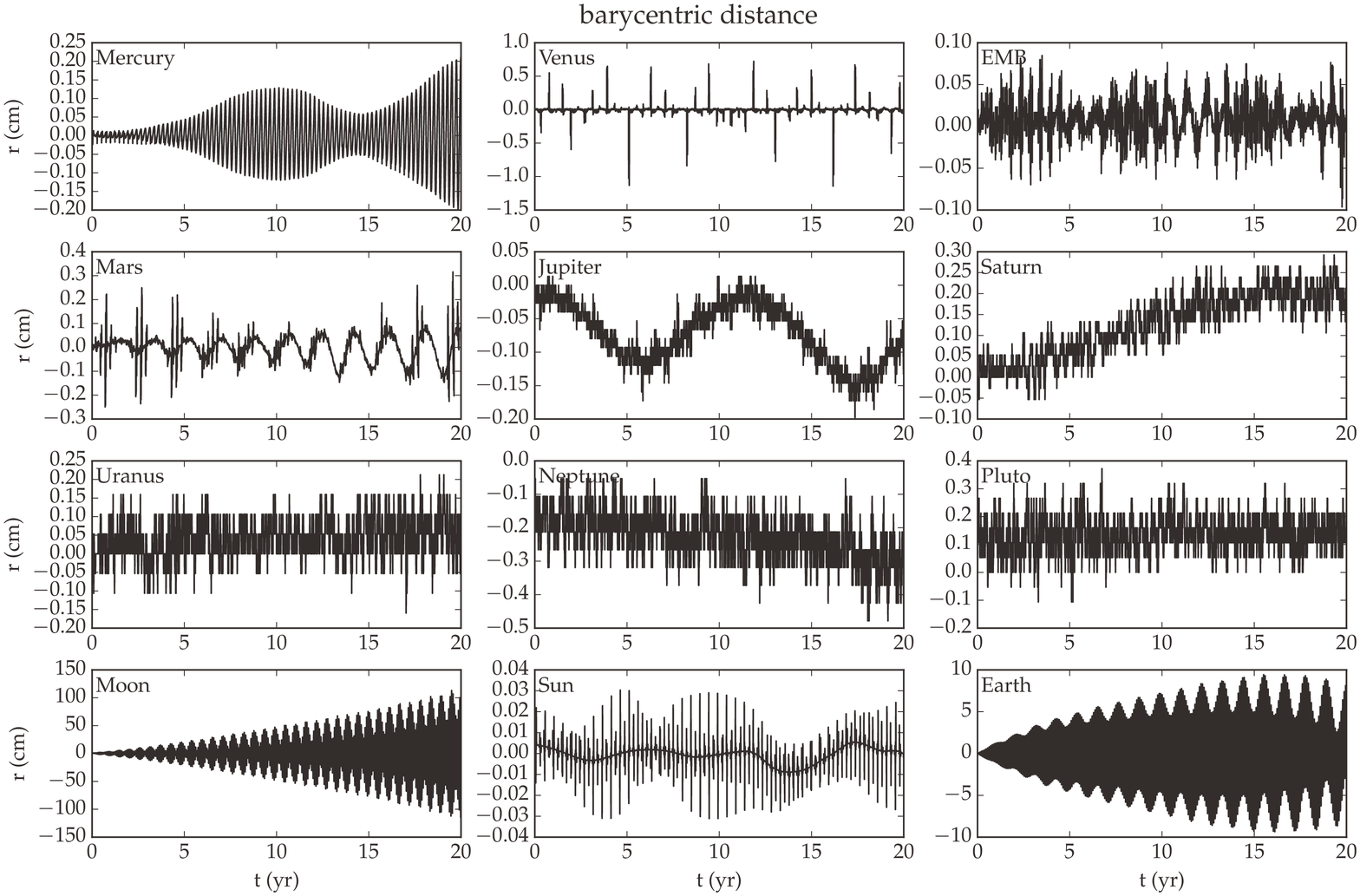}
\includegraphics[width=1.9\columnwidth]{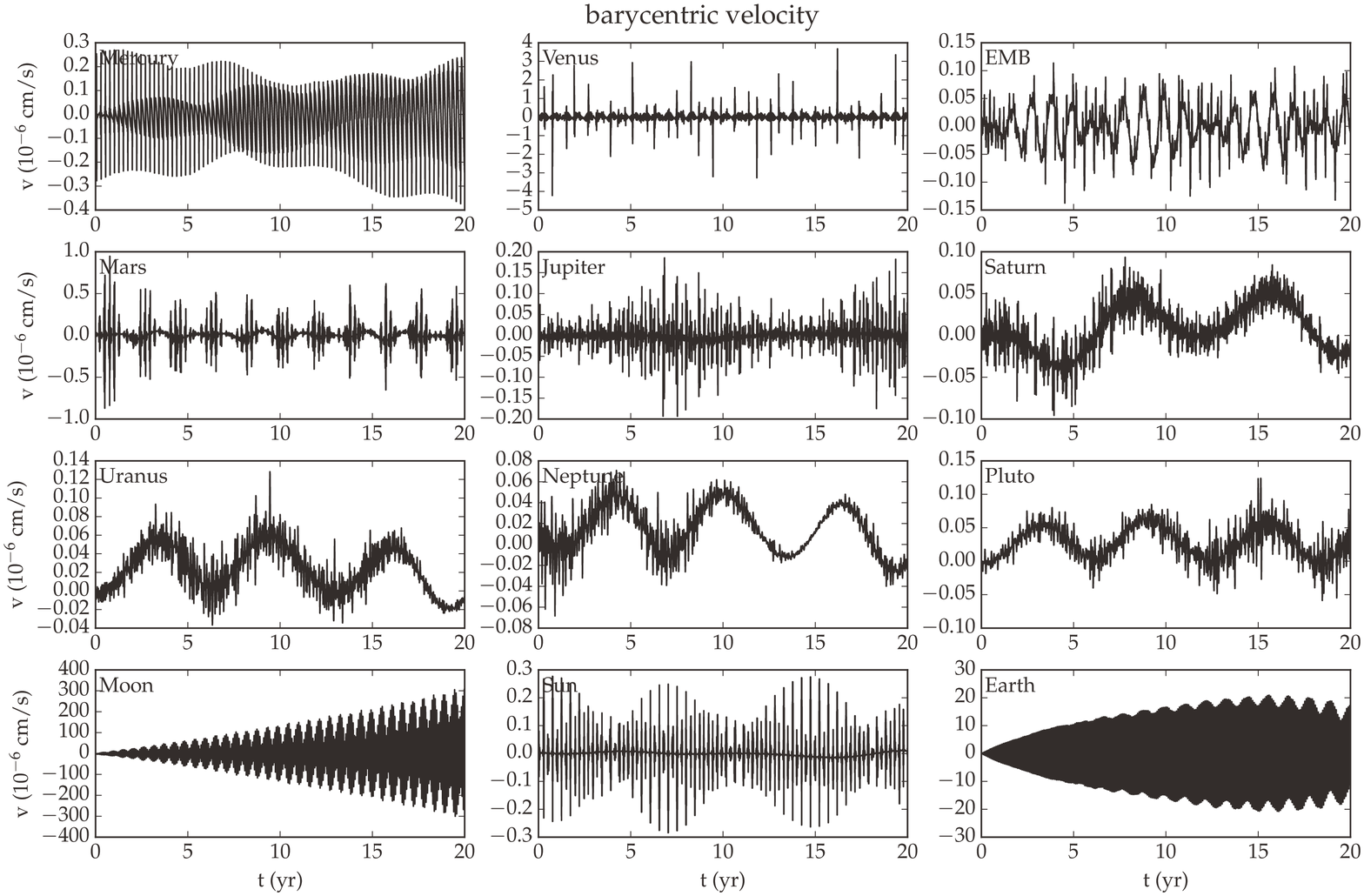}
\caption{Using the same input parameters as DE435, we numerically integrate the \lnm{} model to get a SSE realization and compare it with DE435. This figure shows the differences in the barycentric distances (upper panel) and velocities (bottom panel) of planets (from Mercury to Pluto), with an integration time of 20\,yr. The distance/velocity of the Moon from the Earth-Moon barycentre (EMB), and the barycentric distances/velocities of the Sun and Earth are also compared. The differences in the Moon position and velocity are by far the largest, which results from the details of the modelling of the gravitational interaction between Earth and the Moon.   
}
\label{sse_cp}
\end{center}
\end{figure*}

There are three main ingredients in the SSE creation process, namely the equations of motion which govern the dynamical motions of the bodies, a method for integrating the equations of motion, and the initial conditions (positions and velocities at a given epoch) and dynamical constants such as planetary masses~\citep{Almanac10}.
The initial conditions are determined by fitting numerical integration of the equations of motion to a long history of diverse astronomical observations of the Solar system, such as planetary astrometry,
spacecraft flybys, radar and laser ranging.
The results of the fit are then used with the same equations of motion to predict the positions and velocities of the bodies.

The JPL DE ephemerides are so far the most widely used ephemerides 
in pulsar-timing analyses.
Based on the PMOE ephemeris of Li Guangyu, we make slight modifications to establish 
the \lnm{} dynamical model, which is similar to DE435.
The mathematical framework of PMOE is described in \citet{LGY03} and \citet{LGY08a}, and has been applied to the orbit optimization of LISA~\citep{LGY08b}.
\lnm{} shares similar equations of motion as JPL DE435~\citep{Folkner14}, which includes:
(i) the Newtonian gravity and post-Newtonian corrections between the Sun, Moon, planets and asteroids which are treated as point masses; 
(ii) the figure effects arising from a finite size of the Sun, Earth and Moon;
(iii) the tidal effects of the Earth; and 
(iv) the librations of the Moon.
The numerical integration of the orbits is carried out using the Bulirsch-Stoer method~\citep{NR3}.

For the validation of our study, we demonstrate that we can recover the DE435 solution with the necessary precision.
With the same planetary masses and initial conditions, the result, using a 2-day cadence, 
is compatible with that of DE435 at centimetre level over a 20-yr timespan,
i.e. the difference between the two SSEs can induce residuals of the order of just 1~ns.
The precision of LINIMOSS is therefore sufficient for PTA applications which require $\lesssim$ 100-ns timing precision.
The differences in the barycentric distances and velocities of planets are shown in \FIG{sse_cp}.
This verifies our capabilities in computing planetary orbits and understanding the equations of motion used in DE435.
With the dynamical model we can change a given parameter and numerically integrate to create a modified SSE, so that we can study
the effects of modified planetary masses and orbital parameters on pulsar timing.


\section{Least-square adjustment equation of SSE parameters with pulsar timing}
\label{sec:lsaeq}

\begin{figure*}
\begin{center}
\includegraphics[width=\columnwidth]{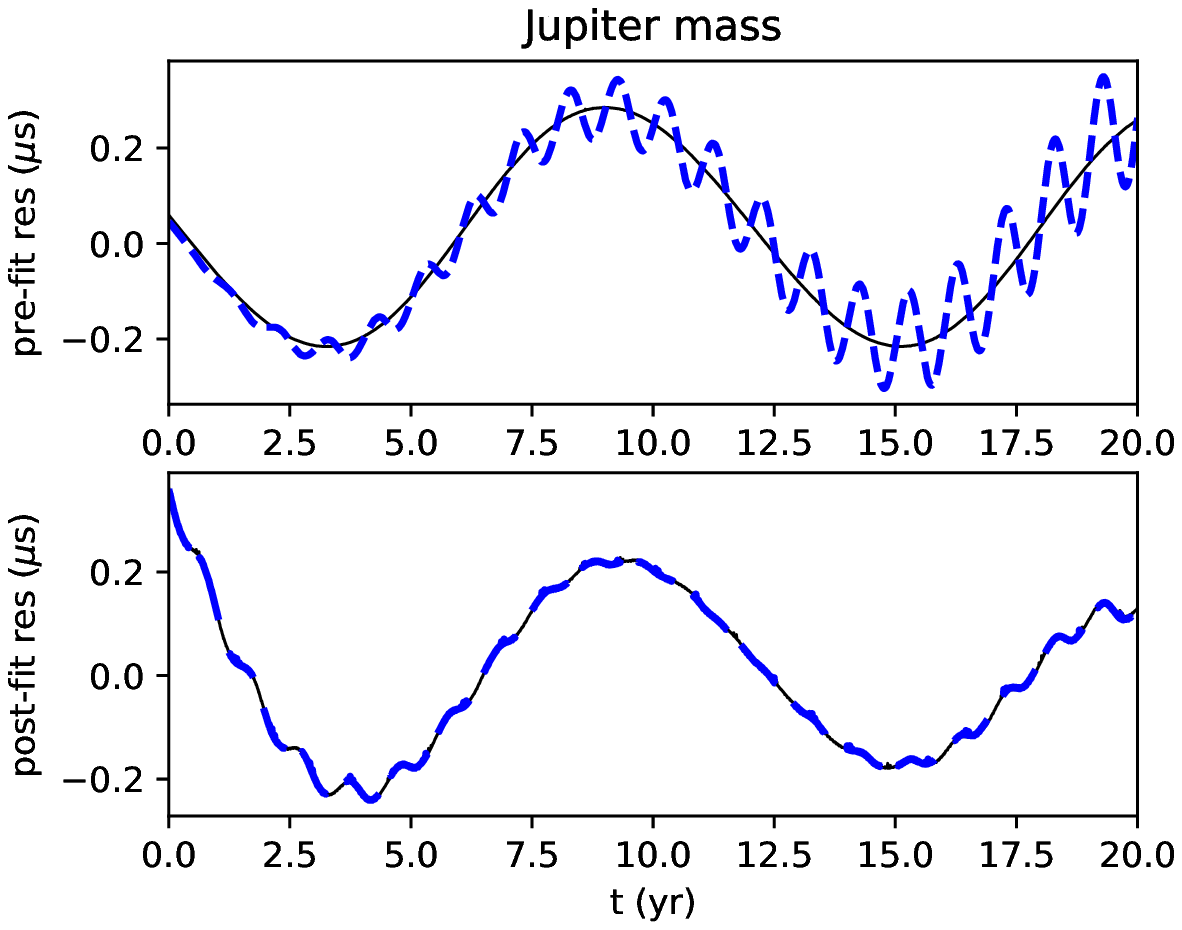}
\includegraphics[width=\columnwidth]{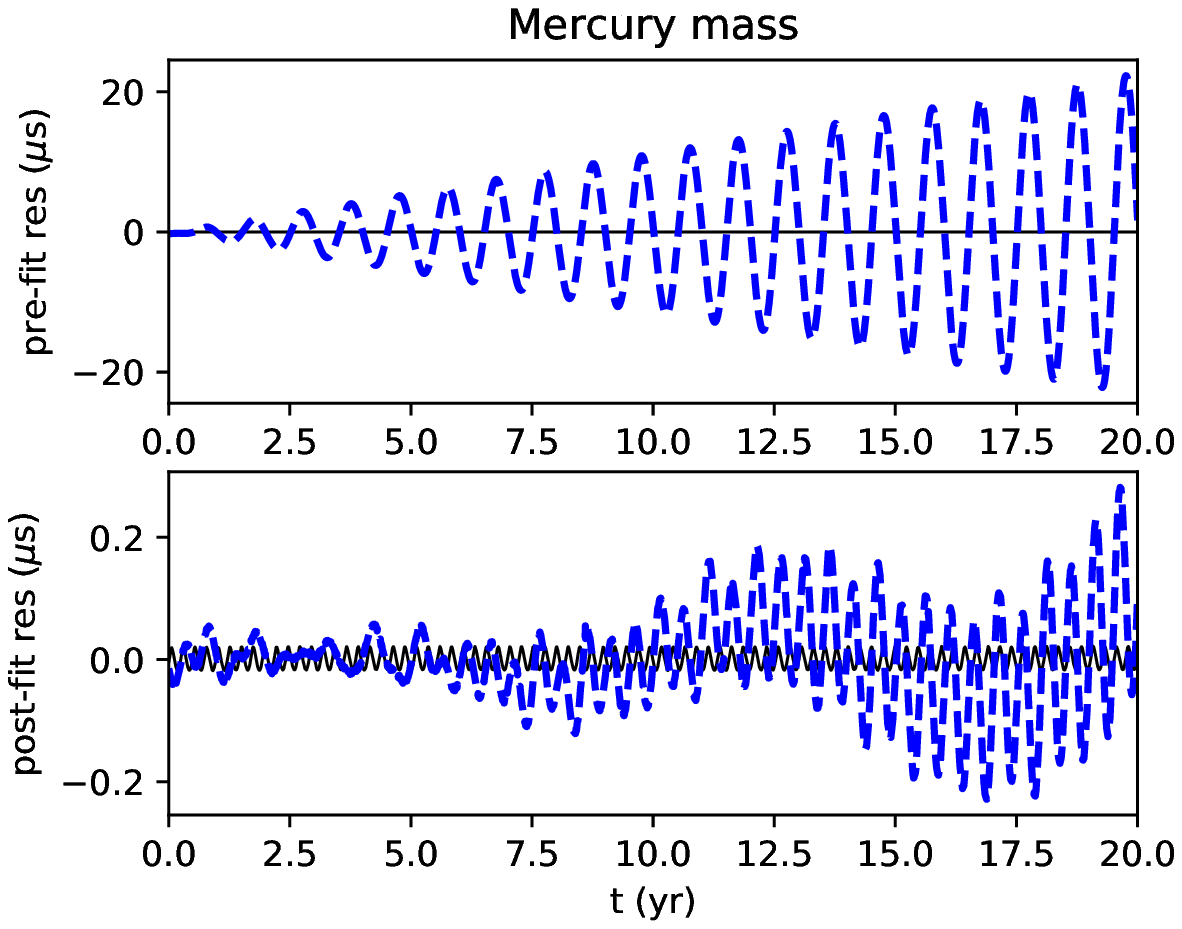}
\caption{The waveform of the timing residuals induced by change of $\delta m=10^{-10}$M$_{\odot}$ in the masses of Jupiter (left column) and Mercury (right column).
The results of \lnm{} dynamical model are in blue, thicker dashed lines, obtained by analyzing the simulated data of PSR J1909$-$3744 using \textsc{tempo2} with the modified SSE.
The waveform of the SSB shift are shown in black, thinner solid lines for comparison, obtained by 
using the ``DMASSPLANET'' parameter of \textsc{tempo2} in the pulsar timing model.
This parameter is designed to model the effects of a modification on the planetary mass while
maintaining the SSE's orbital elements intact, satisfying the assumptions introduced in Champion et al. (2010). 
The top panels show the pre-fit residuals, and the bottom panels are the post-fit residuals after fitting the timing parameters (spin, astrometric and binary parameters). 
Note that the figure aims to show the waveform of timing residuals, so the timing precision of the simulated data are set to be extremely small (0.01\,ns) and the errorbar can be neglected.}

\label{res}
\end{center}
\end{figure*}

Among the terms related to transferring the TOA from the
observatory to the SSB,
the R\o{}mer delay has the largest amplitude,
which is of order $\sim 500$\,s.
This Solar-system R\o{}mer delay is the light's vacuum propagation time between the SSB and the geocentre, determined by the projection of the Earth-SSB vector onto the pulsar direction. 
Therefore, timing residuals induced by SSE errors mainly rely on the change of the barycentric position of the Earth.

Below, we analyze waveforms 
of timing residuals that emulate the effects that possible errors in SSE parameters would cause.
To this end, we use the LINIMOSS dynamical model to create SSEs with perturbations in mass, positions, and velocities of the planets,
and use them in the pulsar timing model to fit simulated data that 
are generated with a reference SSE version. 
The pulsar timing analysis is performed throughout the 
work presented in this paper using \textsc{tempo2} \citep*{HEM06b},
a widely used general-purpose pulsar-timing software package.
The analysis results
formed the basis of a linearized model for Solar-system parameter estimation,
and was then used to explore the effectiveness of previous work and 
demonstrate the improvements we can achieve working with \lnm{}.

\subsection{Methods}

Possible SSE errors are mainly expected as a result of the uncertainties in the masses and dynamical initial conditions of planets.
We use $\VEC \lambda$ to represent 7$\times$10 SSE parameters, i.e. the masses $m$ and initial conditions described by the three-dimensional
vectors of position and velocity, $\VEC{r}$ and $\VEC v$, respectively, of all major planetary systems (in the sequence of 1=Mercury, 
2=Venus, 3=Earth-Moon Barycenter, 4=Moon, 5=Mars, 6=Jupiter, 7=Saturn, 8=Uranus, 9=Neptune and 10=Pluto).
The dependence of the planetary orbits and timing residuals on the SSE parameters $\VEC \lambda$ are non-linear,
but the models can be linearized\footnote{The accelerations in the current dynamical model are analytic functions (in the sense of continuity) with respect to the model parameters. 
We can thus perform linearization, because the velocities and positions of planets, i.e. the time integrals of acceleration, 
must also be analytic \citep{DB11}.} around the reference parameters $\VEC \lambda_{0}$, such that
\begin{equation}
\delta \VEC{t}=\MX D \VEC{(\lambda-\lambda_{0})},
\end{equation}
where $\delta \VEC{t}$ is the column matrix of timing residuals, and $\MX D$ is the design matrix which describes how the timing residuals depend on the SSE parameters.
We test the validity of such linearisation by comparing the waveforms of pulsar timing residuals based on two sets of planet ephemerides, where the perturbation value of first set of ephemerides is given in Appendix A, and the perturbation in the second set are ten times larger. For all parameters, the non-linearity is well bellow 1\% or buried in the noise floor due to the finite numerical precision of \textsc{tempo2},
ensuring in this way that we act well within the linear regime, and that our analysis is self-consistent.

As explained earlier, we take the parameters of the recent DE435 SSE as reference, in order to study the effects on pulsar timing in the presence of small errors in the DE435 parameters.
For all 70 SSE parameters, we perturb one parameter at a time with all others fixed, and integrate the \lnm{} dynamical model to create a modified SSE.
For simplicity, we first try to give the same perturbation magnitude for the parameters of every planet, i.e. $10^{-10}\,$M$_\odot$ for mass, 100\,km for positions and 0.01\,m/s for velocities. 
Because this results in a large diversity in the amplitude of the induced timing residuals, 
we also adjusted the magnitudes of the perturbations so that the induced residuals are in the ns--ms range, 
where the lower limit comes from the numerical precision of \textsc{tempo2}, 
and the upper limit is determined by the spin period magnitude of millisecond pulsars, in order to keep the timing solutions phase-coherent.
We list the parameter perturbations in Appendix A.

Using the \lnm{} dynamical model, we derive the waveform of timing residuals induced by perturbing these SSE parameters.
The process is as follows:
\begin{enumerate}
\item We create 70 modified SSEs, each with one parameter perturbed
\item We simulate TOAs that are predicted by a given timing model and the reference SSE (created by integrating
\lnm{} with the same input parameters as DE435). 
\item These simulated data were then analyzed using tempo2 with the modified SSE. 
\item The timing residuals obtained with the modified SSE are compared to those of the reference SSE to investigate the effect of changing this parameter.
\end{enumerate}
The timing model can be of any MSP in principle, and here we take PSR J1909$-3744$ as an example.
As we focus on deriving the waveform of timing residuals, the simulated timing precision is set to be extremely small (0.01\,ns) to avoid the interference of white noise.
In order to well sample the waveforms and also to probe spectral features at high frequencies, we used a short, 2-day cadence.
We note that while such a cadence is significantly higher 
than for most current individual PTA observation programmes, data sets that result from combinations of data from multiple telescopes and observing campaigns can have an average cadence of the level of a few days \citep{VLH16}. 
However, real PTA data are characterised by irregular cadence
and different cadences may change how power
at various spectral frequencies is absorbed by timing-model parameters.
Special care should be taken when dealing with real data with
regards to such effects.

\subsection{Waveforms and spectra}
For the modified SSE, the barycentric positions of the Earth and other planets will change, which affects the TOAs of a pulsar through various terms including the R\o{}mer delay, timing parallax, gravitational redshift, Shapiro delay and so on \citep*[see e.g.][for details on implementation of such effects in timing algorithms]{EHM06}.
All these effects have been taken into account as we use \textsc{tempo2} to obtain the timing residuals based on a modified SSE.
The major contributions can be understood analytically.
The leading term is 
the variation of the Solar-system R\o{}mer delay, i.e. the light travel time between the geocentre and SSB,
\begin{equation}
\delta t_{1}=-\delta \VEC{r} \cdot \VEC{n}_{\rm p}/c, 
\end{equation}
where $\delta \VEC{r}$ is the difference between the new barycentric position of the Earth (the perturbed Earth orbit relative to the new SSB) and the original position, $\VEC{n}_{\rm p}$ is the 
pulsar's unit barycentric position vector 
and $c$ is the vacuum speed of light.
The variation of the binary R\o{}mer delay, 
the excess light travel time from the pulsar to the binary's barycentre,
will have a small contribution for binary pulsars, with amplitude of
\begin{equation}
\delta t_{2} \sim a_{\rm b}\omega_{\rm b} \delta t_{1}/c,
\end{equation}
where $a_{\rm b}$ is the projected semi-major axis of the pulsar's orbit, and $\omega_{\rm b}$ is the orbital angular frequency.
All other effects are much smaller and can be neglected. 

As discussed above, the change of timing residuals mainly comes from the difference of the barycentric position of the Earth.
The dominant factor that induces $\delta \VEC{r}$ is different when perturbing the masses or orbital parameters of the inner or the outer planets.
For the outer planets $\delta\VEC{r}$ is mainly due to the shift of SSB, but for the inner planets $\delta \VEC{r}$ is dominated by changes in the Earth's orbit.
Using simulations of PSR J1909$-$3744, \FIG{res} shows the pre-fit timing residuals for variations in the masses of Mercury and Jupiter, as well as the post-fit residuals after fitting the standard timing parameters (including spin, astrometric and binary parameters). 

The signals introduced by the perturbations in the outer-planets parameters are easier to understand, and they are roughly consistent with previous work. 
For an increase in the mass of Jupiter by $10^{-10}$M$_{\odot}$, the resulting pre-fit residuals 
include a sinusoid at Jupiter's orbital period together with a smaller annual term with increasing amplitude.
The Jupiter term comes from the displacement of the SSB, while the annual term results from a slight variation in the orbital period of the Earth.
The small annual term becomes negligible after fitting the timing model (its power absorbed by timing astrometric parameters), 
as shown in the bottom left panel of \FIG{res}. 
Previous works only model the SSB displacement with Jupiter's period, but neglects the annual term.
Our result for the outer planets is similar to the tests reported by \citet{Champion10}, 
where they created a new ephemeris that had identical parameters to DE421, except for a small decrease (of 7$\times10^{-11}$M$_{\odot}$) in the mass of Jupiter.

However, the differences between the results of previous work and the dynamical model
become significant for the inner planets, 
where additional complexity arises because of their lager effects on the orbit of the Earth-Moon system. 
For a change of $10^{-10}$M$_{\odot}$ in the mass of Mercury, the pre-fit residuals are dominated by the annual term, 
which is much larger than the sinusoid at Mercury's orbital period.
Fitting the timing model absorbs the annual term, but a signal at 0.5-yr remains, 
with an amplitude still larger than that of the Mercury term, breaking down the basic assumption of previous work.
The signal with half-year period comes from the change in the eccentricity of Earth orbit. 
One may expect that the fit of the timing parallax will absorb this signal. 
The waveform of timing parallax also has a half-year period, and the phase is $2(\lambda_{p}-\lambda_{E})$, where $\lambda_{p}$ and $\lambda_{E}$ are the ecliptic latitudes of the pulsar and Earth respectively. 
The phase cannot be adjusted arbitrarily, as it is mainly determined by fitting the pulsar position to the annual term. 
So fitting for parallax does not mitigate the half-year signals with different phases.

\begin{figure*}
\begin{center}
\includegraphics[width=2\columnwidth]{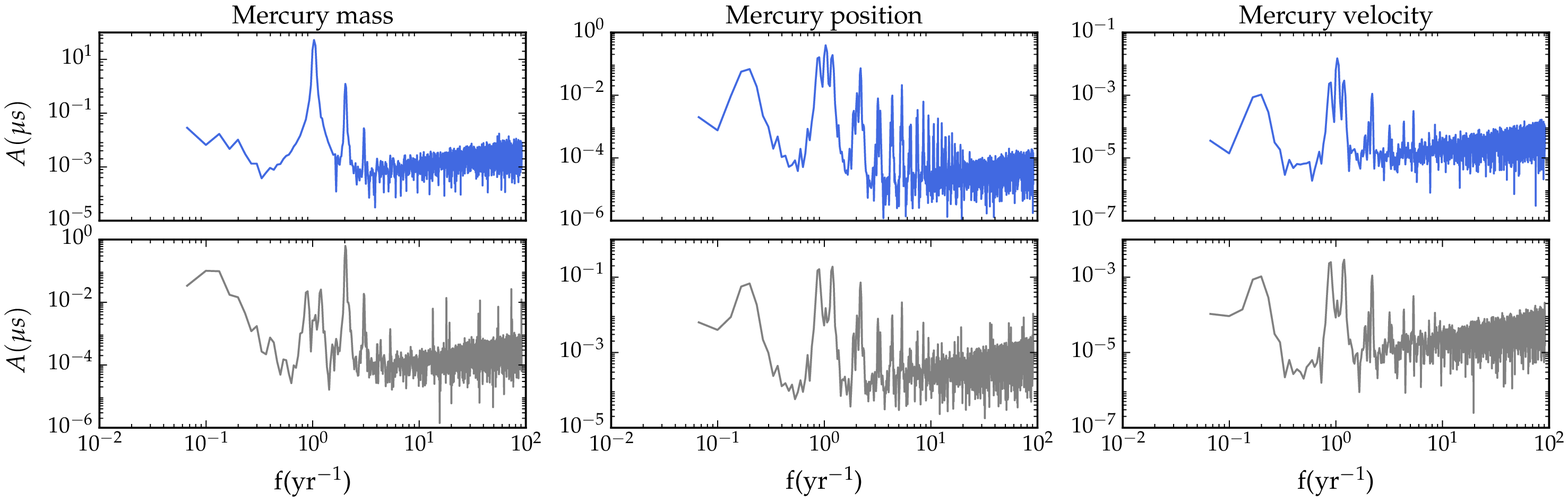}
\includegraphics[width=2\columnwidth]{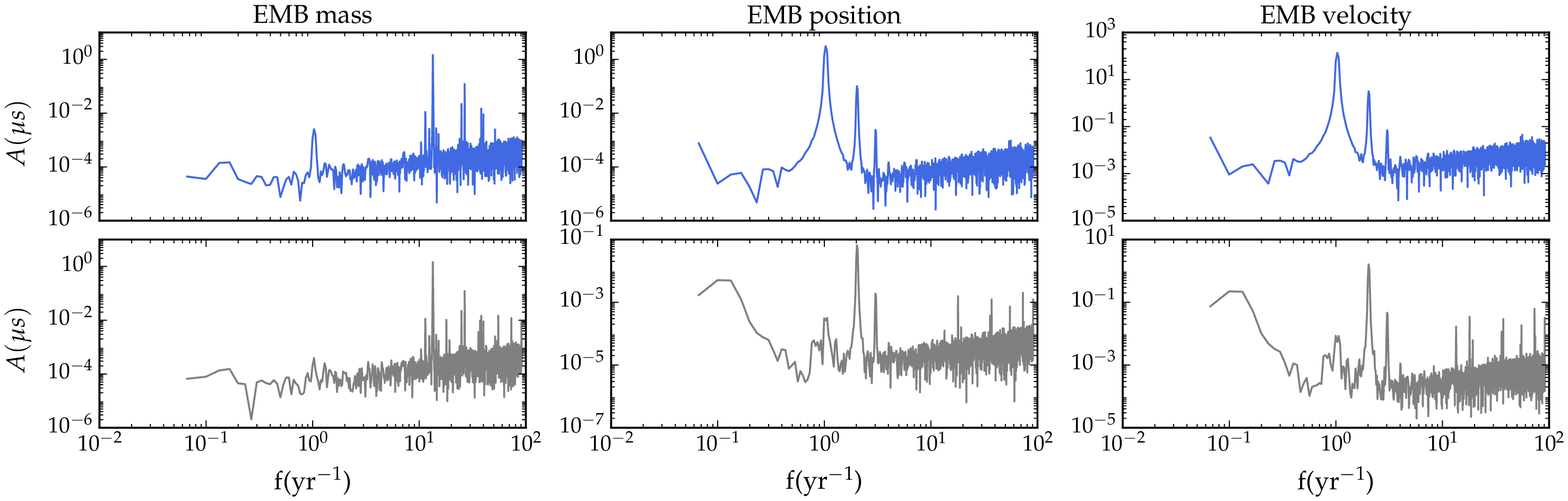}
\includegraphics[width=2\columnwidth]{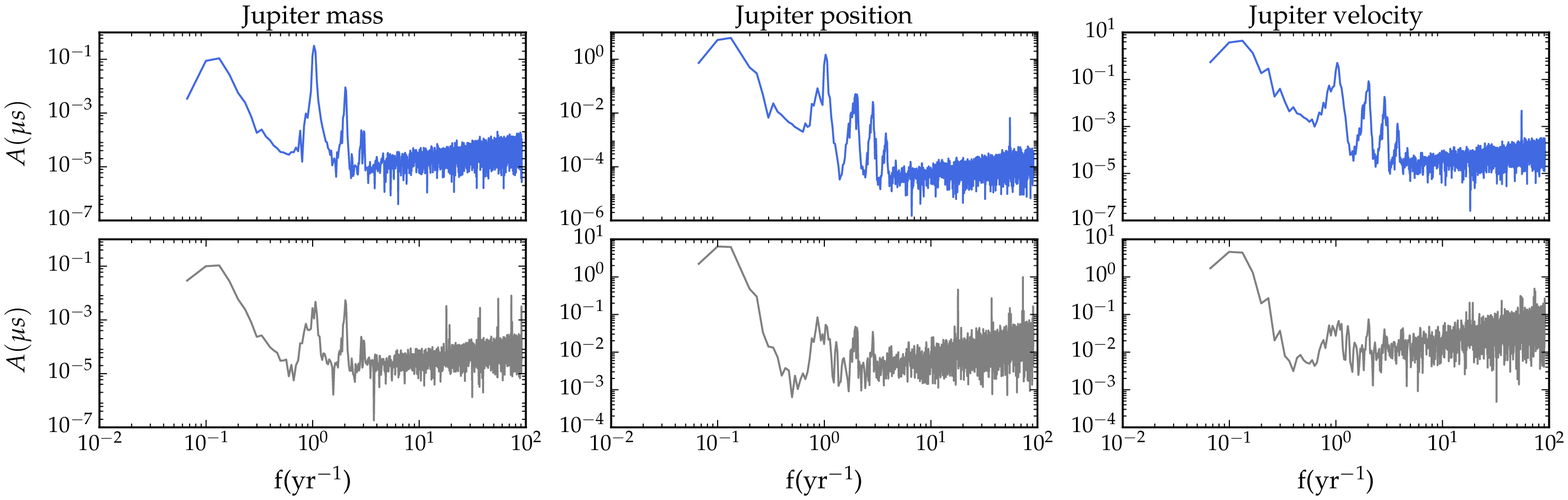}
\includegraphics[width=2\columnwidth]{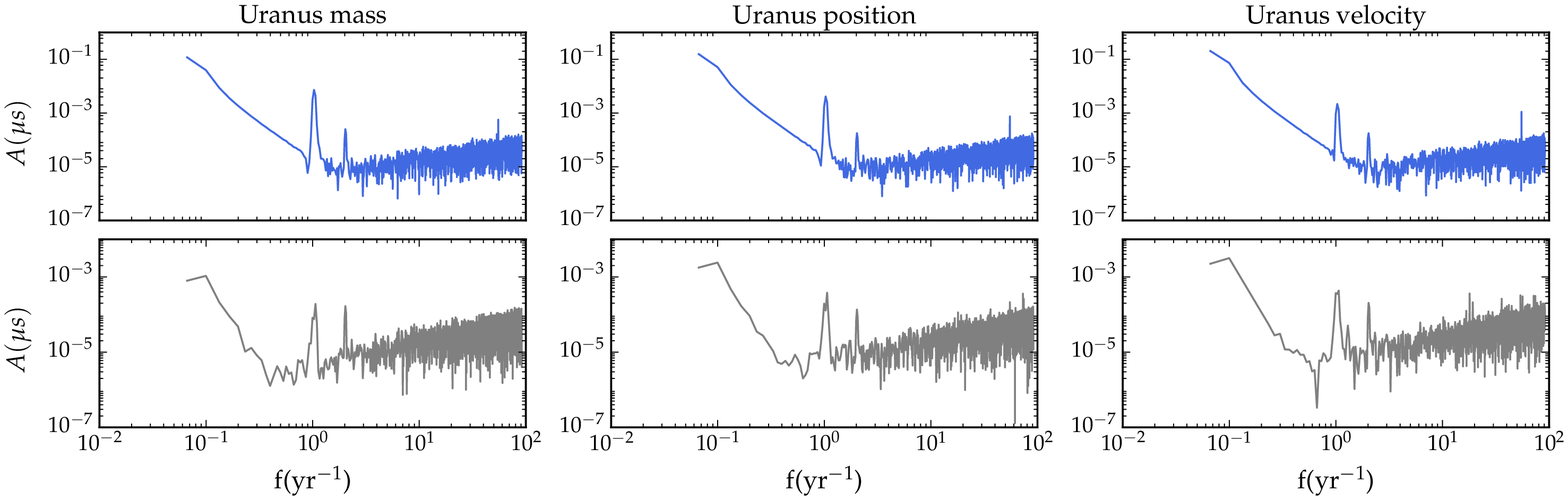}
\caption{The power spectra of the timing residuals induced by perturbing the SSE parameters. We show the cases of Mercury, EMB, Jupiter and Uranus. 
For every object, the three columns correspond (from left to right) to the change of mass, position and velocity, and the two rows show the pre-fit (top/blue colour) and post-fit (bottom/grey colour) results.}
\label{fft}
\end{center}
\end{figure*}

We perform spectral analysis of the timing residuals caused by perturbing each SSE parameter using Fourier transform and the Hann window function.
In order to more clearly probe the spectral characteristics at Jupiter's orbital period, we extended the data timespan to 30\,yr.
The power-spectral density can be written in the form~\citep{LBJ14}, 
\begin{equation}
S(f)=\frac{A(f)^{2}}{f},
\end{equation}
such that $A(f)$, in unit of time, is the characteristic amplitude, i.e. the amplitude of the component at frequency $f$.
The benefit of using such a specification is that $A(f)$ has the dimension of time regardless of the spectrum's steepness.
The power spectra of pre-fit and post-fit timing residuals are shown in \FIG{fft} for four chosen cases, i.e. Mercury (inner planet), EMB (directly change the Earth parameter), Jupiter (outer planet with orbital period shorter than the data span) and Uranus (outer planet with orbital period longer than the data span).
We give here a simple qualitative discussion of the spectra (including the ones not shown in \FIG{fft}).
Note that fully understanding the behavior is complex and beyond the scope of this paper.
\begin{enumerate}
{\bf \item Inner planets:} For the inner planets, the power spectra are dominated by frequencies from the change of the Earth's orbit, together with many other frequency components. 
The common frequencies correspond to periods of 1\,yr and 0.5\,yr, presumably caused by variations in the orbital period and eccentricity of the Earth (for a detailed explanation, see Appendix B).
There are specific frequencies for each planet, such as $\sim$5\,yr for Mercury and $\sim$8\,yr for Venus.
These values coincide with the resonance periods of a system composed of the perturbed planet and Earth, 
which are close to the common multiples of the two planets' periods\footnote{Similar phenomena have been reported by \citet{IF99}, where they calculate the time ephemeris TE405 (transformation from terrestrial time TT to barycentric coordinate time TCB) based on DE405. They compare the difference between TE200 and TE405, and find that the largest terms are 4 sinusoids with the Uranus and Neptune synodic periods (relative to Earth) and sidereal periods, as the Uranus and Neptune masses have the largest relative change from DE200 to DE405 of all the mass parameters of the JPL ephemerides.}.
{\bf \item Earth-Moon system.} We also perturb the parameters of the Earth-Moon system itself.
It was expected that the mass error of Earth cannot be detected with pulsar timing, because the one-year signal will be absorbed by fitting pulsar position.
However, perturbing the Earth-Moon system also leads to secondary effects in addition to the annual term, such as signals with periods of 0.5 year and 1 month.
The signal with 1-month period comes from the perturbation of the Moon, 
and would not be found if cadence of several weeks were to be used.
Here we simulate with fine sampling frequency to get the full characteristics of the spectra.
Besides, perturbing the Earth parameter is much more effective on changing timing residuals than the parameters of other planets, since the latter only influences the Earth orbit indirectly.
With the same magnitude of perturbation, the signal introduced by Earth is still much larger than that of other planets even after fitting timing parameters.
{\bf \item Outer planets:} For the outer planets, the power spectra consist of the orbital periods of the planet and Earth, as shown by the case of Jupiter.
The situation is slightly different when the orbital period of the planet is longer than the data span, such as the case of Uranus.
The orbital frequency is so small that it is beyond the frequency range of the spectra.
In this case there is no peak at the planet period in the spectra of pre-fit residuals, but only a plateau at the low-frequency end.
From Jupiter to Pluto, the Earth term decreases as the planet is farther away from Earth.
\end{enumerate}

The power spectra of the post-fit residuals are also shown in \FIG{fft}.
The SSE-related residuals can be fitted out by timing parameters only at certain frequencies.
The annual term will be absorbed by fitting the position and proper motion of the pulsar.
The spin period and period derivative of the pulsar will absorb the 
low-frequency (frequencies below $1/T$, where $T$ the dataset's timespan) components\footnote{ Note that, as discussed in \cite{CGL18}, when pulsar noise contains low-frequency (time-correlated) noise,
	this noise component and low-frequency signals from outer planets with orbital periods longer
	that the data length are covariant, and appropriate low-frequency noise modelling is important for properly assessing the uncertainties when
	estimating mass and orbital parameter values for these cases.}.
However, the signals cannot be completely mitigated by fitting the timing parameters, with the power at other frequencies still left.
So appropriate dynamical modelling is necessary to accurately account for the SSE errors. 



\begin{table}
\caption{The uncertainties of SSE parameters constrained by pulsar timing, assuming 30-yr observations of 3 pulsars with timing precision of 100\,ns and cadence of 2 weeks. The position and velocity uncertainties
refer to the total uncertainty, i.e. the square root of the sum of variances 
for each of the x, y, z axes.}
\begin{center}
\begin{tabular}{|c|l|l|l|}
\hline
\hline
Planet & $\delta m\,($M$_\odot)$ & $\delta r\,$(km) & $\delta v\,$(m/s) \\
\hline
Mercury & $ 3.9 \times 10^{ -11 }$  & $ 2.5 \times 10^{ 3 }$  & $ 4.0 \times 10^{ -1 }$  \\
Venus & $ 6.8 \times 10^{ -13 }$  & $ 6.1 \times 10^{ 1 }$  & $ 2.0 \times 10^{ -2 }$  \\
EMB & $ 2.3 \times 10^{ -13 }$  & $ 2.6 \times 10^{ -1 }$  & $ 3.1 \times 10^{ -5 }$  \\
Moon & $ 1.2 \times 10^{ -14 }$  & $ 1.7 \times 10^{ -1 }$  & $ 3.3 \times 10^{ -4 }$  \\
Mars & $ 7.2 \times 10^{ -13 }$  & $ 9.7 \times 10^{ 2 }$  & $ 1.1 \times 10^{ -1 }$  \\
Jupiter & $ 4.4 \times 10^{ -11 }$  & $ 6.5 \times 10^{ 1 }$  & $ 1.0 \times 10^{ -3 }$  \\
Saturn & $ 2.6 \times 10^{ -10 }$  & $ 2.0 \times 10^{ 3 }$  & $ 1.4 \times 10^{ -2 }$  \\
Uranus & $ 2.3 \times 10^{ -9 }$  & $ 4.0 \times 10^{ 4 }$  & $ 2.8 \times 10^{ -1 }$  \\
Neptune & $ 2.3 \times 10^{ -9 }$  & $ 4.0 \times 10^{ 4 }$  & $ 3.9 \times 10^{ -1 }$  \\
Pluto & $ 2.3 \times 10^{ -9 }$  & $ 4.1 \times 10^{ 8 }$  & $ 3.7 \times 10^{ 3 }$  \\
\hline
\end{tabular}
\end{center}
\label{par_err}
\end{table}%


\begin{figure*}
\begin{center}
\includegraphics[width=2\columnwidth]{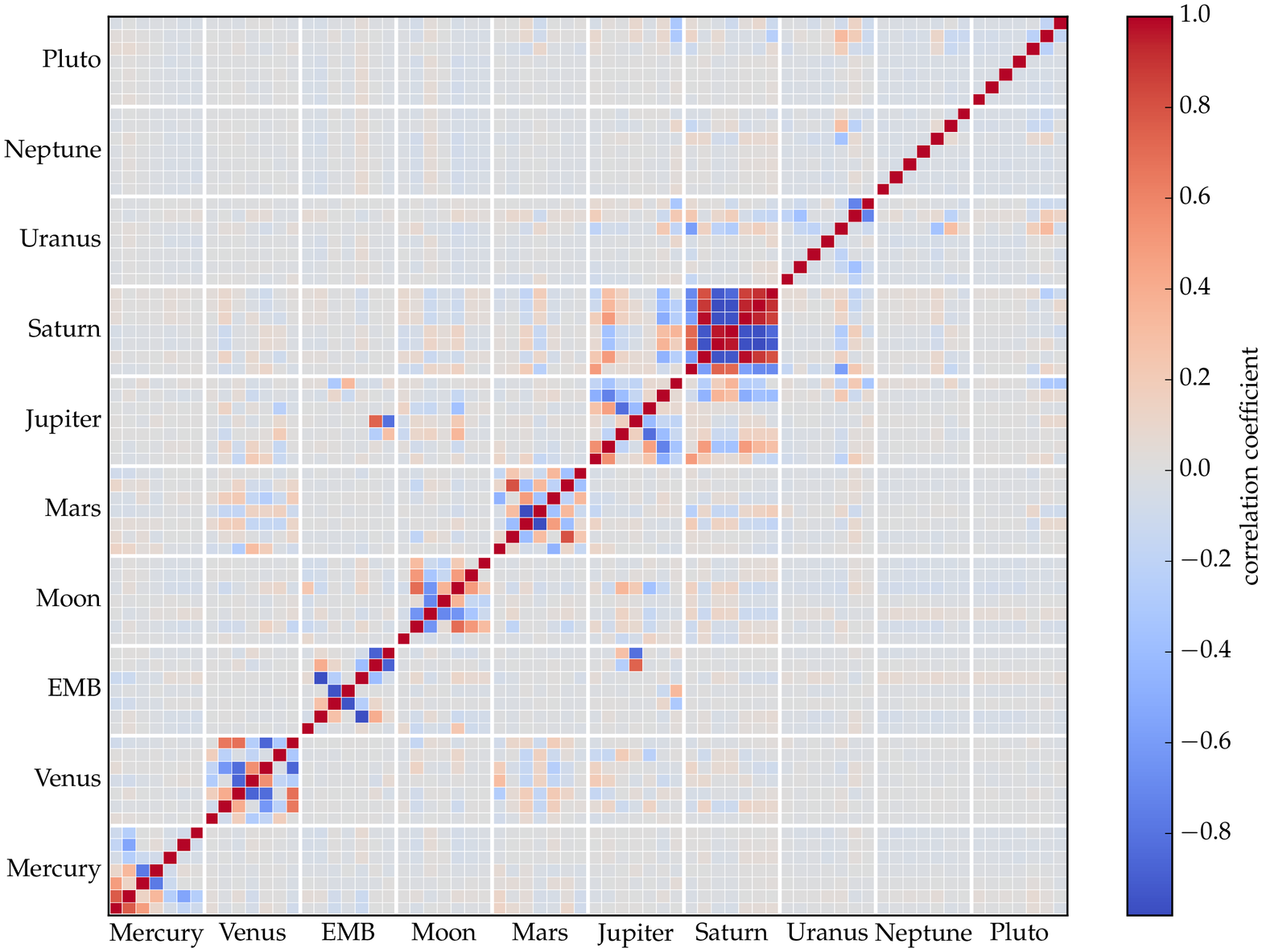}
\caption{The correlation coefficients of SSE parameters, in the sequence of mass, 3-vector position and 3-vector velocity of Mercury, Venus, EMB, Moon, Mars, Jupiter, Saturn, Uranus, Neptune, Pluto. The color indicates the value of the correlation coefficient, which is defined as $\langle \delta \lambda_i \delta \lambda_j \rangle/ \sqrt{\langle \delta \lambda_i ^2\rangle \langle \delta \lambda_j^2 \rangle}$.}
\label{corr}
\end{center}
\end{figure*}

\subsection{Parameter estimation}
The deviations of SSE parameters from the underlying true values will influence the pulsar TOAs and lead to specific timing signals, as discussed above.
Thus, pulsar timing can be used to constrain these SSE parameters by fitting the SSE-related signals in the data.
Assuming that the stochastic noise in the timing residuals follows a Gaussian distribution, the accuracy achievable in measuring the parameters $\VEC\lambda$ can be estimated using the Cram\'er-Rao bound~\citep[see e.g.][]{Lee16},
\begin{equation}
	\left \langle \delta \lambda_{i} \delta \lambda_{j}\right\rangle 
	=\left(\MX{D}^\MX{T} \MX{C}^{-1} \MX{D} \right)_{i,j}^{-1}\,.
\label{CRB}
\end{equation}
where the $i$ and $j$ indices denote pairs of parameters, $\MX D$ is the design matrix, and $\MX{C}$ is the covariance matrix of the timing noise.
We combine all SSE parameters and timing parameters to form the design matrix, such that the correlation between these parameters can be taken into account.
With 30 yr of biweekly timing observations of 3 pulsars with 100\,ns precision (assuming only white, time-uncorrelated pulsar noise), 
the constraints on the SSE parameters are calculated according to \EQ{CRB} and the values are listed in \TAB{par_err}. 

Since the signal of an inner planet is much larger than the SSB shift, 
one may expect that the planetary mass can be better constrained with full dynamical modelling than previous work.
However, the signals of some SSE parameters are highly correlated, 
as pulsar timing is mainly relevant to the barycentric position of the Earth. 
\FIG{corr} shows the correlation matrix of the SSE parameters.
The uncertainty in \TAB{par_err} is derived by simultaneously fitting all the SSE parameters.
This will be larger than the uncertainty obtained by fitting one parameter only, which is mainly determined by the signal amplitude.
In \cite{Champion10}, the secondary effects of varying Jupiter mass is tested using full dynamics. 
However, apart from the mass of Jupiter, the rest of the mass and orbital
parameters were held fixed, and it was thus unclear how the different SSE parameters correlate in parameter inference.
Detailed study is needed to analyze which parameter is well constrained by other observations, such that the correlation can partly be broken.
If one can put constraints on some parameters using independent observations, the uncertainties could be effectively reduced.


\section{Conclusion}
\label{sec:end}
We have developed and implemented the \lnm{} dynamical model of the Solar system based on the PMOE ephemeris, 
which is adequate for pulsar-timing applications, with differences from DE435 at centimeter level over a 20\,yr timespan.
We investigate the timing residuals induced by uncertainties on the SSE parameters.
For the outer planets the SSE-related timing residuals are dominated by the shift of the SSB, 
and generally consistent with previous work, which assumes
the orbits fixed to the SSE input values. 
For the inner planets, the variation of the Earth's orbit is more prominent than the displacement of SSB, 
making the approach in previous work sub-optimal. 
As noted also in \cite{CGL18}, the approach in previous work poses limitations that especially for the 
inner planets require further investigations due to the high precision of new pulsar-timing data. 
By using a fully dynamical analysis, we are able to additionally take into account 
the effects of errors in the orbital elements and the covariance between the parameters we attempt to measure,
making our search for errors in SSE parameters and planetary mass constraints more robust.
We also show that the frequency components of the SSE-related signals are complicated, 
which are not completely absorbed by fitting the timing parameters and may as a consequence
complicate searches for correlated signals in PTA data.
Particularly, one can see complex line-type spectra, which may cause problem for GW search using PTAs.
The spikes can potentially introduce false signal in the search of single source.
Therefore, fully dynamical modelling is needed to appropriately deal with the SSE errors.
This subject will be discussed in a separate paper.
We also estimate the uncertainties of the SSE parameters constrained by pulsar timing
and present the correlation matrix, noting that covariances between SSE parameters
are expected to pose limitations in the precision with which we will be able to limit their values.

\section*{Acknowledgments}
This work was supported by XDB23010200, National Basic Research Program
of China, 973 Program, 2015CB857101 and NSFC U15311243, 11690024, 11373011. 
We are also supported by the MPG funding for the Max-Planck Partner Group. 
G. Y. Li led this work for some time.
His unfortunate passing-away makes it impossible for him
to be the corresponding author,
but we acknowledge his valuable and pioneering work related to this paper.




\bibliographystyle{mn2e}
\bibliography{ms} 



\appendix

\section{List of parameter perturbations}

We list the details of SSE parameter perturbations we used for this study.
The x, y, z index for the position and velocity indicates that the same 
perturbation value was used for all three axes.

\begin{tabular}{cccc}

\hline
\hline
Planet & $\delta m\,($M$_\odot)$ & $\delta r_{\textrm{x,y,z}}\,$(km) & $\delta v_{\textrm{x,y,z}}\,$(m/s) \\
\hline
Mercury & $10^{-11}$  & $10^{2}$  & $10^{-2}$  \\
Venus & $10^{-11}$  & $10^{2}$  & $10^{-2}$  \\
EMB & $10^{-14}$  & $10^{-2}$  & $10^{-6}$  \\
Moon & $10^{-14}$  & $10^{-2}$  & $10^{-5}$  \\
Mars & $10^{-11}$  & $10^{2}$  & $10^{-2}$  \\
Jupiter & $10^{-10}$  & $10^{2}$  & $10^{-3}$  \\
Saturn & $10^{-10}$  & $10^{2}$  & $10^{-3}$  \\
Uranus & $10^{-10}$  & $10^{3}$  & $10^{-2}$  \\
Neptune & $10^{-10}$  & $10^{3}$ & $10^{-2}$  \\
Pluto & $10^{-10}$  & $10^{7}$  & $10^{2}$  \\
\hline

\end{tabular}

\section{Simple analysis of elliptical orbits}

The timing signals induced by perturbing the SSE parameters are complicated, with many components.
It is hard to quantitatively predict the signal analytically,
but we can gain some insight into the main features qualitatively by analyzing the simple case of an elliptical orbit with small eccentricity.
We discuss the terms with one-year and half-year period when we change the mass of inner planets.
Note that we only give a possible explanation of the spectra characteristics. Fully understanding the spectra is beyond the scope of this paper.

Variation in the mass of inner planets changes both the orbital period and eccentricity of the Earth.
Considering the simplified case that the Earth orbit is a elliptical and unperturbed,
the orbital frequency $f$ is determined by Kepler's third law, 
\begin{equation}
f=\frac{1}{2\pi}\sqrt{\frac{GM}{a^{3}}},
\end{equation}
where $G$ is the gravitational constant, $M$ is the equivalent mass inside the Earth orbit, and $a$ is the semi-major axis of the orbit.
An increase in $M$ will change the orbital frequency.
The eccentricity is related to the total energy $E$ by
\begin{equation}
E=\frac{GM(e^{2}-1)}{2a}.
\end{equation}
An increase in $M$ will change the potential energy without affecting the kinetic energy, so the eccentricity will also change.

Change in the orbital period will lead to a one-year signal, which is easy to understand.
Considering the simplest case of a circular orbit, when the orbital frequency increases by $\delta f$, the variation in the orbit is (ignoring the zero point of phase),
\begin{equation}
\delta x= a \sin (2\pi ft) - a \sin (2\pi (f+\delta f) t),
\end{equation}
which can be expanded as 
\begin{equation}
\delta x 
= -2 a \cos \left(2\pi (f+\frac{\delta f}{2}) t\right)  \sin \left(2\pi \frac{\delta f}{2}t\right) 
\label{dxdf}
\end{equation}
This is a sinusoid with (higher) frequency $f+\delta f/2\approx 1\,{\rm yr}^{-1}$ modulated by a sinusoid with a very low frequency of $\delta f/2$.
When $\delta f$ is very small, the second sinusoid approximates $\pi \delta ft$ within the data span of pulsar timing, and \EQ{dxdf} becomes
\begin{equation}
\delta x \approx -2\pi a \, \delta f\,t \cos \left(2\pi(f+\frac{\delta f}{2}) t\right)
\end{equation}
This is an annual term with amplitude growing with time.

Change in the eccentricity will lead to a signal with half-year period.
The position vector in the orbital plane can be expressed in terms of the eccentric anomaly $u$,
\begin{eqnarray}
x&=&a(\cos u-e), \\
y&=&a\sqrt{1-e^{2}}\sin u.\\
z&=&0，
\end{eqnarray}
We take the x component, for instance, to analyze the change of the orbit caused by
a change in eccentricity, $\delta e$, and the derivation for the y component is similar.
The partial derivative of $x$ with respect to $e$ is
\begin{equation}
\frac{\partial x}{\partial e}= -a\sin u \frac{\partial u}{\partial e} -a  . \label{pxpe}
\end{equation}
According to Kepler's equation,
\begin{equation}
u+e\sin u=\frac{t}{f},
\end{equation}
$\delta u$ and $\delta e$ are related by
\begin{equation}
\delta u + \delta e\sin u + e\cos u \, \delta u =0 .
\end{equation}
Thus we have
\begin{equation}
\frac{\partial u}{\partial e}=-\frac{\sin u}{1+e\cos u} . \label{pupe}
\end{equation}
Inserting \EQ{pupe} into \EQ{pxpe} gives
\begin{equation}
\frac{\partial x}{\partial e}=  \frac{a \sin^{2} u}{1+e\cos u} -a 
\end{equation}
When the eccentricity is very small, we have 
\begin{equation}
\delta x \approx \left( -\frac{1}{2}a\cos (2u)  - \frac{1}{2}a \right) \delta e.
\end{equation}
So the half-year signal may result from the change of Earth's eccentricity.


\bsp	
\label{lastpage}
\end{document}